\documentclass{ws-ijgmmp}

\begin{document}

\markboth{Borowiec, God{\l}owski, Szyd{\l}owski}
{Dark matter and dark energy........}

\catchline{}{}{}{}{}
%
%%%%%%%%%%%%%%%%%%%%%%%%%%%%%%%%%%%%%%%%%%%%%%%%%%%%%%%%%%%%%%%%%%%%

\title{Dark matter and dark energy as a effects of Modified Gravity
\footnote{}}

\author{Andrzej Borowiec}

\address{
Institute of Theoretical Physics, University of Wroc{\l}aw\\
 pl. Maksa Borna 9, 50-204  Wroc{\l}aw, Poland\,}
%\email{borow@ift.uni.wroc.pl\footnote{}} }

\author{W{\l}odzimierz God{\l}owski}

\address{
Astronomical Observatory  Jagiellonian University\\
30-244 Krakow, ul. Orla 171, Poland\,}
%\email{godlows@oa.uj.edu.pl\_name\footnote{}} }

\author{Marek Szyd{\l}owski}

\address{
Astronomical Observatory  Jagiellonian University\\
30-244 Krakow, ul. Orla 171, Poland\,\\
Mark Kac Complex  System Research Centre \\
Jagiellonian University,  30-059 Krakow, ul. Reymonta 4, Poland\,}

%\email{szydlo@oa.uj.edu.pl\_name\footnote{}} }

\maketitle

\begin{history}
\received{(Day Month Year)}
\revised{(Day Month Year)}
\end{history}

\begin{abstract}
We explain the effect of dark matter (flat rotation curve) using
modified gravitational dynamics. We investigate in this context a
low energy limit of generalized general relativity with a nonlinear
Lagrangian ${\cal L}\propto R^n$, where $R$ is the (generalized) Ricci scalar 
and $n$ is parameter estimated from SNIa data. We estimate parameter
$\beta$ in modified gravitational potential $V(r) \propto
-\frac{1}{r}(1+(\frac{r}{r_c})^{\beta})$. Then we compare value of
$\beta$ obtained from SNIa data with $\beta$ parameter evaluated
from  the best fitted rotation curve. We find $\beta \simeq 0.7$
which becomes in good agreement with an observation of spiral galaxies
rotation curve. We also find preferred value of $\Omega_{m,0}$
from the combined analysis of supernovae data and baryon oscillation
peak. We argue that although amount of "dark energy" (of
non-substantial origin) is consistent with SNIa data and flat curves
of spiral galaxies are reproduces in the framework of modified
Einstein's equation we still need substantial dark matter. For
comparison predictions of the model with predictions of  the
$\Lambda$CDM concordance model we apply the Akaike and Bayesian
information criteria of model selection.
\end{abstract}

\keywords{Dark Energy; Dark Matter; Modified Gravity.}

\section{Introduction}

Different astronomical observations
\cite{Riess:1998cb,Perlmutter:1998np} are pointing out that our
Universe becomes, at present time, in accelerating phase of expansion.
In  principle, there are two quite different approaches to
explain this observational fact. In the first approach (which
can be called substantial) it is assumed that universe is filled by
mysterious perfect fluid violating the strong energy condition 
$\rho_x+3p_x>0$, where $\rho_x$ and $p_x$ are, respectively, the energy 
density and the pressure of this fluid. The
nature as well as origin of this matter, called dark energy, is
unknown until now. Among these approaches have appeared concordance
$\Lambda$CDM model, which predicts that baryons contribute only
about 4\% of the critical  energy density, non-baryonic cold dark
matter (CDM) about 25\% and the cosmological constant $\Lambda$
(vacuum energy) remaining 70\%. Although $\Lambda$CDM model fits well SNIa data
\cite{Riess:2004nr,Astier:2005} this model offers only description of
the observations not  their explanation. From the methodological view point
the conception of mysterious dark energy seems to be 
effective physical theory only and motivates theorists for searching of
alternative approaches in which nature of dark energy will be known
at the very beginning.

In the first approach it is assumed that Einstein's theory of general
relativity is valid which reduces in practice  (after assuming 
 Robertson-Walker symmetry of space slices) to the case of 
 Friedman - Robertson - Walker models. Nevertheless, theoretically
it is not a'priori excluded the possibility of cosmology based on  some
extension of Einstein's general relativity. In this paper we consider 
such particular cases.

On the other hand, there are alternative ideas of explanation, in
which instead of dark energy some modifications of Friedmann's
equation are proposed at the very beginning. In these approaches
some effects arising from new physics like brane cosmologies,
quantum effects, nonhomogeneities effects etc. can mimic dark energy by a
modification of Friedmann equation. Freese \& Lewis
\cite{Freese:2002sq} have shown that contributions of type $\rho^n$
to Friedmann's equation $3H^2=\rho_{\mathrm{eff}}$, where
$\rho_{\mathrm{eff}}$ is the  effective energy density and $n$ is a
constant, may describe such situations phenomenologically. These
models (called by their authors called the Cardassian models) give
rise to acceleration, although the universe is flat, contains the
usual matter and radiation without any dark energy components. This
models have been tested by many authors (see for example
\cite{Sen03,Dev03,Z1,Z2,Z3,Z4,Godl04}). 
What is still lacking is a fundamental theory
(like general relativity) from which these models can be derived
after postulating Robertson Walker symmetry.

In this paper we shall consider such particular type of generalization of
Einstein's general relativity in which Lagrangian is proportional to
$R^n$, where $R$ is generalized Ricci skalar. In particular, Einstein's 
general relativity is recovered if we put $n=1$. This theory is part of
the larger class of so-called f(R) gravity, i.e. theories derived from
gravitational Lagrangians that are analytical (usually polynomial)
functions of $R$. (see e.g. \cite{Cap02,Carroll04,Nojiri03,Flanagan03,Allemandi04,Mota,Mena,Clifton,Cruz06}. 
In this approach (we called it non-substantial) instead of postulating
mysterious dark energy it is assumed some extension of general relativity.
Then effect of acceleration appeared naturally as a
dynamical effect of the model. For modified  gravity
one can find Newtonian potential in non-relativistic limit and ask about
possibility to explain flat rotation curve of spiral galaxies -
major evidence for dark matter in the universe
\cite{Stelle78,Milgrom83,Cap06}. 
(However,  see also \cite{Salucci} for non flat rotational curves.)
The main goal here is to
explore power of this particular generalization of gravity
in the context of dark energy and dark matter problems. We
argue that although cosmology with modified Lagrangian
${\cal L}\propto R^n$ can explain "dark energy problem" but baryon
oscillation test distinguishes value of density parameter of 
mater to be equal $\Omega_{m,0} \simeq 0.3$, i.e. problem of dark matter is yet 
not solved within the framework of f(R) theories. We also demonstrate
that models under considerations can reproduce rotation curves of
spiral galaxies.

The structure of the paper is as follows. In section II we
define class of cosmological models of essential theory of gravity
with lagrangian proportional to the Ricci scalar. Section III is
devoted to analyze constraints on model parameters from SNIa, baryon
oscillation peak and CMB shift. In section IV we investigate
problem of rotation curves of spiral galaxies. Section V 
summarizes our results and formulates general conclusion that
models modified gravity which are based on generalized lagrangians
${\cal L}\propto R^n$ and Palatini formalism although solve the
acceleration and flat rotation curves problems still favor
$\Omega_{m,0} \simeq 0.3$.

\section{Cosmological models of nonlinear Palatini gravity}

Let us consider the simplest cosmological model of generalized
Einstein's theory of gravity with Lagrangian  ${\cal L}=f(R)$ which
is function of the (generalized) Ricci scalar. The action is assumed in 
the form
\begin{equation}
A=A_{\mathrm{grav}}+A_{\mathrm{mat}}=\int (\sqrt{\det g}f(R)+2L_{\mathrm{mat}} 
(\Psi) )d^{4}x
\end{equation}
where $f(R) \propto R^n$ and $n$ is a constant.  We also assumed that 
dynamical equation determining evolution of the cosmological model 
can be derived from the action through the Palatini formalism in which 
both metric $g$ and symmetric connection $\Gamma$ are regarded as an independent variables. Thus $R\equiv R(g, \Gamma)=g^{\mu\nu}R_{\mu\nu}(\Gamma)$ denotes generalized Ricci scalar (see e.g. \cite{Allemandi04} for details).
 
Because of homogeneity and isotropy of the surface $t=const$ is assumed,
$t$ - being a  global cosmic time, we choose Robertson-Walker metric i.e. 
\begin{equation}
d s^2=-d t^2+a^2 (t) \Big[ {1 \over {1-k r^2}} d r^2+ r^2 \Big( d \theta^2 +\sin^2 (\theta) d \varphi^2  \Big) \Big]\ .
\label{RW1}
\end{equation}
where $k=0,\pm1$ is curvature index, $r,\theta,\phi$ are usual
spherical coordinates. 
Some properties of these theories (in the Palatini framework)  have been already
investigated by Capozziello et al. \cite{Cap06c}. It has been
demonstrated that under two popular choices $f(R) \propto R^n$ and
$f(R) \propto \ln R$ both models provide well fits to the SNIa data.

Here we consider matter content in the form of perfect fluid which satisfy the
conservation condition:
\begin{equation}
\frac{d \rho}{d t}= -3H(\rho+p)
\end{equation}
where $H=\frac{d}{d t} (\ln a)$ is Hubble'a function. For convenience 
we assume simple form of equation of state (E.Q.S). After adopting
Palatini formalism field equation reduces %for the models with R-W symmetry 
to the ordinary second order differential equation which admits first 
integral in the form:
\begin{equation}
\label{eq:xh}
H^2=F(a).
\end{equation}
The first integral (\ref{eq:xh}) is usually called (generalized) Friedman equation.
This is the first order  differential equation in which right hand side
is determined by matter content and curvature. Due to simple relation between
the scale factor and the redshift $z$ ($1+z=\frac{a_0}{a}$) formula (\ref{eq:xh})
can be written in the form $H^2=F(z)$ ($a_0$ denotes the present value of the scale
factor which corresponds to the  redshift $z=0$)
In the system filled by both dust matter and radiation
the function $F(a)$ takes the following form:
\begin{eqnarray}
H^2 = \frac{2n}{3(3-n)}
\left[\frac{\kappa \eta_{dust}}{\beta} \right]^{\frac{1}{n}}\,a^{-\frac{3}{n}}+
\nonumber \\
+\frac{4n(2-n)\kappa\eta_{rad}}{3\beta(n-3)^2}
\left[ \frac{\kappa \eta_{dust}}{\beta} \right]^{\frac{1-n}{n}}\,a^{-\frac{n+3}{n}}
- \frac{k}{a^2}\, \left[ \frac{2n}{(n-3)} \right]^2 \ .
\label{eq:MFRn}
\end{eqnarray}
where $\mathcal{L}_{\mathrm{grav}}=\frac{\beta}{2-n}R^n \sqrt{g}$,
$\beta$ is dimensional constant, $w_{dust} \equiv p_{dust}/\rho_{dust} =0$.
%(or $TrT \ne 0$). 
If $n=\beta=1$ and $\eta_{rad}=0$ then the classical
FRW dust filled model is recovered.

Let us formulate some important remarks:
\begin{enumerate}
\item the formula (\ref{eq:MFRn}) contains many nonphysical parameters which
can be replaced by dimensionless density parameters $\Omega_i$ defined for 
each additive contribution to the r.h.s. of (\ref{eq:xh}). This in turn
can be treated as a (fictitious or real) component of some effective
energy density. Density parameter $\Omega_i$ is defined for each
energy component in standard way $\Omega_i=\rho_i/ 3H_0^2$ where
$H_0$ is present value of the Hubble function and $\rho_i$ is energy
density of $i-th$ fluid.
\item for our further analysis 
it is useful to separate this contribution on the r.h.s. of (\ref{eq:xh})
which represents real dust matter scaling like $a^{-3}$ from
the non-substantial effects of generalized Lagrangian (related
with n-parameter). Then our basic formula can be rewritten to the
new more suitable form:
\begin{eqnarray}
\left(\frac{H}{H_0}\right)^2=
\Omega_{m,0}(1+z)^{3}\frac{2n}{\left(3-n\right)}
\Omega_{nonl,0}(1+z)^{\frac{3\left(1-n\right)}{n}}+ \nonumber \\
+\Omega_{r,0}(1+z)^{4}\frac{4n\left(2-n\right)}{\left(n-3\right)^2}
\Omega_{nonl,0}(1+z)^{\frac{3\left(1-n\right)}{n}} \ .
\label{eq:7}
\end{eqnarray}
where parameter $\Omega_{nonl,0}$ is determined  from the
constraint $H(z=0)=H_{0}$%, which can be easily reduces to:
\begin{equation}
\label{eq:8}
\Omega_{nonl,0}=\left(\frac{2n}{\left(3-n\right)}\Omega_{m,0}+
\frac{4n\left(2-n\right)}{\left(n-3\right)^2}\Omega_{r,0}\right)^{-1} \ .
\end{equation}
Here $k=0$ is assumed for simplicity (for more general formulas see \cite{Bor06}).
\end{enumerate}

One can check that in the case of $n=1$ one obtains Einstein de Sitter
model filled with matter and radiation. The basic formula (\ref{eq:7})
will be suitable 
in the next section for providing priors on $\Omega_{m,0}$ which can be 
obtained from independent extragalactic measurements or baryon oscillation peak.

From formula (\ref{eq:7}) on can drive a few conclusions. The first is that
rejection of $\Omega_{r,0}$ in (\ref{eq:7}) doesn't eliminate
automatically the dust term.
On the other hand substitution $\eta_{dust}=0$ give rise to
rejection of the second (radiation) term automatically. The next observation
arising from (\ref{eq:7}) is that term
$\Omega_{nonl,0}(1+z)^{\frac{3\left(1-n\right)}{n}}$ plays a role
of lapse function. Therefore one can re-scale original cosmological
time following the rule: $t\mapsto \tau$:
$d\tau^2=\Omega_{nonl,0}(1+z)^{\frac{3\left(1-n\right)}{n}} dt^2$
and then obtain, after re-scaling density parameters,  
a flat model which is dynamically equivalent to the flat FRW
model with : $\overline{\Omega_{m,0}}=\Omega_{m,0} \frac{2n}{3-n}$ and $\overline{\Omega_{r,0}}=\Omega_{r,0}
\frac{4n\left(2-n\right)}{\left(3-n\right)^2}$. Therefore, the exact solutions 
are well known in the form of
$t=t(\overline{\Omega_{m,0}},\overline{\Omega_{r,0}},z)$.

It is also worth to notice that equation (\ref{eq:7}) is equivalent to
\begin{equation}
\label{eq:9}
\left(\frac{H}{H_0}\right)^2-F(a) \equiv 0 \qquad \mathrm{or} \qquad
\frac{\left(a^{'}\right)^{2}}{2}+ V(a)=0,
\end{equation}
where $'\equiv \frac{d}{d\tau}$, $dt|H_0|=d\tau$,
$V(a)= -\rho_{eff} \frac{a^2}{6}= \frac{-F(a)a^2}{2}$.
Due to particle like representation of the dynamics in the form (\ref{eq:9})
it is possible its investigation in terms of what H.-J. Schmidt calls classical
mechanics with the lapse function \cite{Schmidt}.

\section{Observational constraints on  modified gravity parameters}

Within the framework of modified gravity, the
acceleration originates from non-substantial contribution
arising from curvature modification.  This gives rise to negative effective
pressure and leads to self accelerating cosmology.

\subsection{Constraining model parameters from SNIa data}

The fundamental test for  parameters of cosmological model  
is based on the luminosity distance as a function of red-shift $d_l(z)$
(the so-called Hubble diagram)
\begin{equation}
\label{eq:10}
d_L(z) =  (1+z) \frac{c}{H_0} \frac{1}{\sqrt{|\Omega_{k,0}|}}
\mathcal{F} \left( H_0 \sqrt{|\Omega_{k,0}|} \int_0^z \frac{d z'}{H(z')} \right) \ ,
\end{equation}
where $\Omega_{k,0} = - \frac{k}{H_0^2}$ and $\mathcal{F}(x) \equiv
\sinh (x),  x, \sin(x) \qquad \mathrm{for} \qquad  k=-1, 0, +1$
respectively. For distant SNIa relation between luminosity distance
$d_L$, absolute magnitude $M$ and directly observed their apparent
magnitude $m$ has the following form:
\begin{equation}
\label{eq:10b}
\mu \equiv m - M = 5\log_{10}d_{L} + 25= 5\log_{10}D_{L} + \mathcal{M} \ ,
\end{equation}
where $\mathcal{M} = - 5\log_{10}H_{0} + 25$ and $D_{L}=H_{0}d_{L}$.

The goodness of fit is characterized by the parameter
\begin{equation}
\label{eq:10c}
\chi^{2}=\sum_{i}\frac{(\mu_{i}^{\mathrm{theor}}-\mu_{i}^{\mathrm{obs}})^{2}}
{\sigma_{i}^{2}} \ .
\end{equation}
where $\mu_{i}^{\mathrm{obs}}$ is the measured value,
$\mu_{i}^{\mathrm{theor}}$ is the value calculated in the model under
consideration, and $\sigma_{i}$ is the total measurement error.
Assuming that supernovae
measurements come with uncorrelated Gaussian errors, one can determine the
likelihood function
$\mathcal{L}\propto \exp(-\chi^{2}/2)$.
The Probability Density Function (PDF) of cosmological parameters
\cite{Riess:1998cb} can be derived from Bayes' theorem. Therefore, one can
estimate model parameters by using a minimization procedure. It is based on
the likelihood function as well as on the best fit method  minimizing $\chi^2$.

In our analysis  we used two samples of supernovae. One of them
is ``Gold'' Riess et al. sample of 157 SNIa \cite{Riess:2004nr}.
Second one is the sample of 115 supernovae compiled recently by
Astier et al. \cite{Astier:2005}. This latest sample of 115
supernovae is our basic sample.

For statistical analysis we have restricted the parameter
$\Omega_{m,0}$ to the interval $[0,1]$ and $n$ to $[-10.0,10.0]$
(except $n=0$ and additionally $n=3$ for $w=0$). Moreover, because
of the singularity at  $n=3, w=0$ we have separated the cases $n>3$
and $n<3$ for $w=0$ in our analysis. Please note that
$\Omega_{nonl,0}$ is obtained from the constraint $H(z=0)=H_{0}$.
The results of two fitting procedures performed on Riess and Astier
samples  with different prior assumptions for $n$ are presented in
Tables 1 and 2. In the Table 1 the values of model parameters
obtained from the minimum the $\chi^2$ are given, whereas in Table 2
the results from marginalized probability density functions are
displayed. The best fit (minimum $\chi^2$) gives $n \simeq 2.6$ with
the Astier et al. sample versus $n \simeq 2.1$ with the Gold sample.
In Figure 1 we present Probability distribution  obtained  with the
Astier sample for the parameters $\Omega_{m,0}$ and $n$ for
non-linear gravity model, (case $n<3$ marginalized
 over the rest of parameters). Please note that from Fig. 1 we
obtain a very weak dependence of PDF on the matter density parameter if only
$\Omega_{m,0} \ge 0.05$. Because bounce is a generic features of presented
models for $n>2$ \cite{Bor06} it is interesting to calculate from observational 
data probability that value of $n$ paprameter is greated from two. We find that
$P(n>2) \simeq 0.99$. It means that bounce is strongly favored over big-bang 
scenario like to in loop quantum gravity for example \cite{szyd06}.
The Fig. 2 and Fig. 3 shows likelihood contours on the  plane $(\Omega_{m,0},n)$ obtained
(from fits to the SNIa data and baryon oscilation peak test respecitvely), 
obtained for non-linear gravity model, for the case $n<3$ marginalized 
over ${\cal M}$

\begin{table}
\caption{The flat non-linear gravity model with dust and radiation.
Results of statistical analysis performed with Astier et al. and
Gold Riess et al. samples of SNIa as a minimum $\chi^2$ best-fit. We
separately analyzed the case $n>3$ and $n<3$ .} \label{tab:1}
\begin{tabular}{@{}p{1.5cm}rrrr}
\hline  \hline
sample & $\Omega_{m,0}$ & $\Omega_{nonl,0}$ & $n$ &  $\chi^2$ \\
\hline
%\startdata
Gold   &  0.35 &$<0.01$ & 3.001&180.7     \\
$n<3$  &  0.89 &$ 0.23$ & 2.13 &181.5     \\
$n>3$  &  0.35 &$<0.01$ & 3.001&180.7     \\
\hline
Astier &  0.01 &$-1.47$ & 3.11 &108.7     \\
$n<3$  &  0.98 &$ 0.08$ & 2.59 &108.9     \\
$n>3$  &  0.01 &$-1.47$ & 3.11 &108.7     \\
\hline
%\enddata
\end{tabular}
\end{table}

\begin{table}
\caption{The flat non-linear gravity cosmological model with dust and
radiation. The values of the parameters obtained  from
one dimensional PDFs calculated on the Astier et al. and the
 Gold Riess et al. SNIa samples. Because of the singularity
at $n=3$ we separately analyze the cases $n>3$ and $n<3$ .}
\label{tab:2}
\begin{tabular}{@{}p{1.5cm}rrr}
\hline \hline
sample & $\Omega_{m,0}$ & $\Omega_{nonl,0}$ & $n$  \\
\hline
Gold  &$ 0.01^{}$ & $0.26^{}_{}$ & $2.11^{}_{}$ \\
$n<3$ &$ 1.00_{}$ & $0.26^{}_{}$ & $2.11^{}_{}$ \\
$n>3$ &$ 0.01^{}$ & $-0.01^{}_{}$& $3.001^{}_{}$\\
\hline
Astier&$ 0.01^{}$ & $0.09^{}_{}$ & $2.56^{}_{}$ \\
$n<3$ &$ 1.00^{}$ & $0.09^{}_{}$ & $2.56^{}_{}$ \\
$n>3$ &$ 0.01^{}$ &-$0.01^{}_{}$ & $3.01^{}_{}$ \\
\end{tabular}
\end{table}

\begin{figure*}[ht!]
\includegraphics[width=0.45\textwidth]{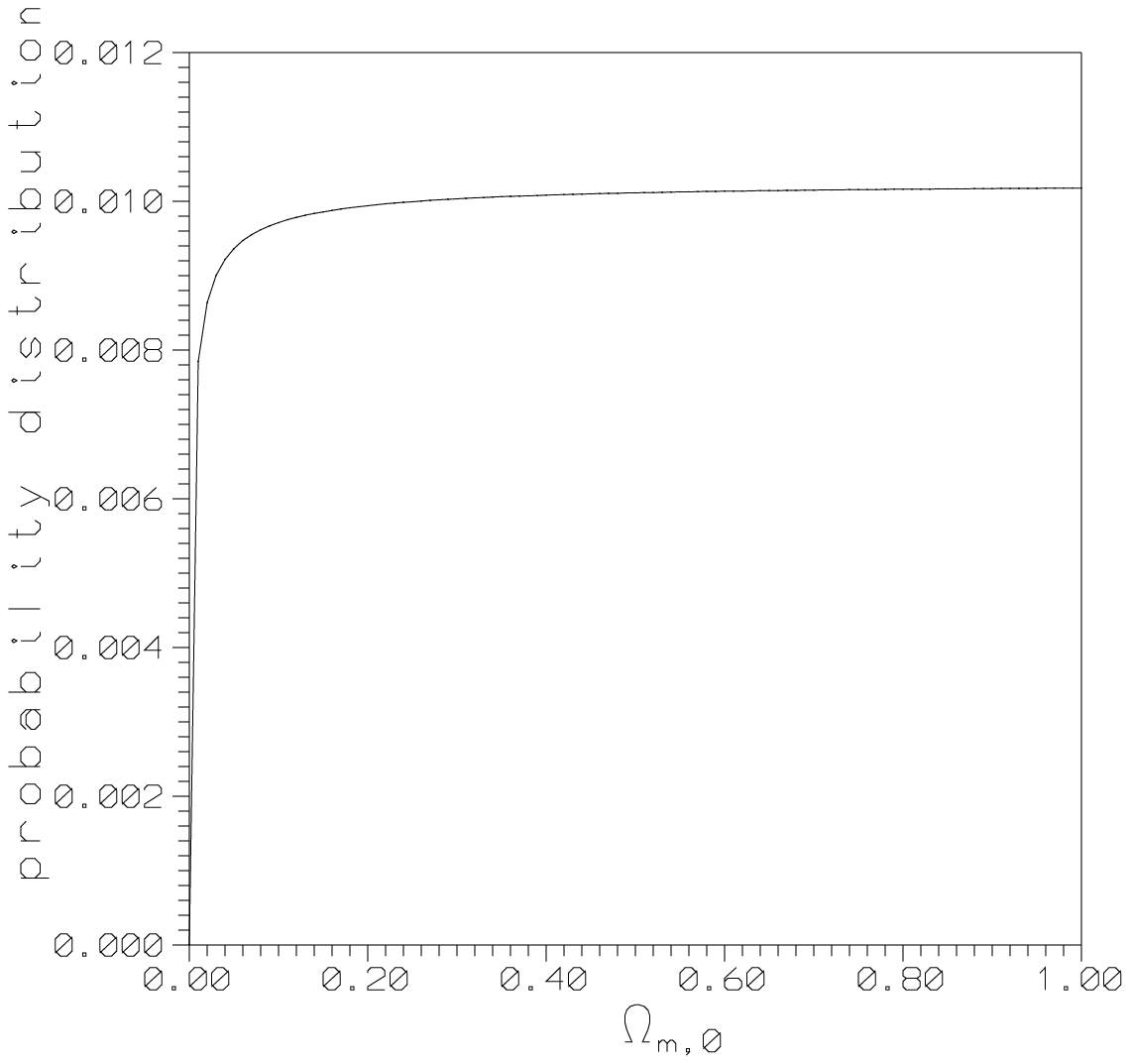}
\includegraphics[width=0.45\textwidth]{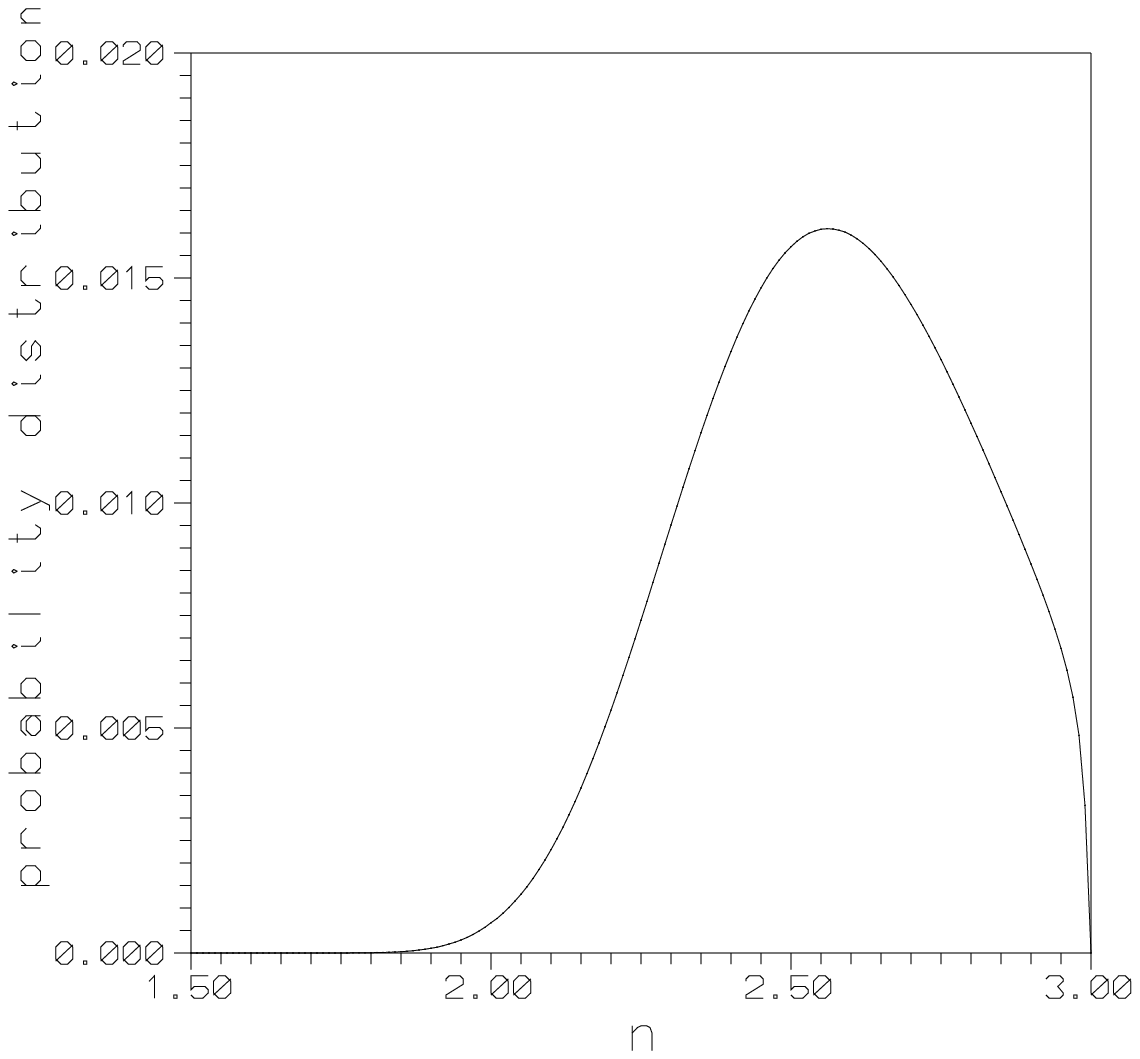}
\caption{Probability distribution  obtained  with the Astier sample
for the parameters $\Omega_{m,0}$ and $n$, marginalized over the
rest of parameters. Non-linear gravity model, $n<3$ . }
\label{fig:1}
\end{figure*}

\begin{figure*}[ht!]
\includegraphics[width=0.9\textwidth]{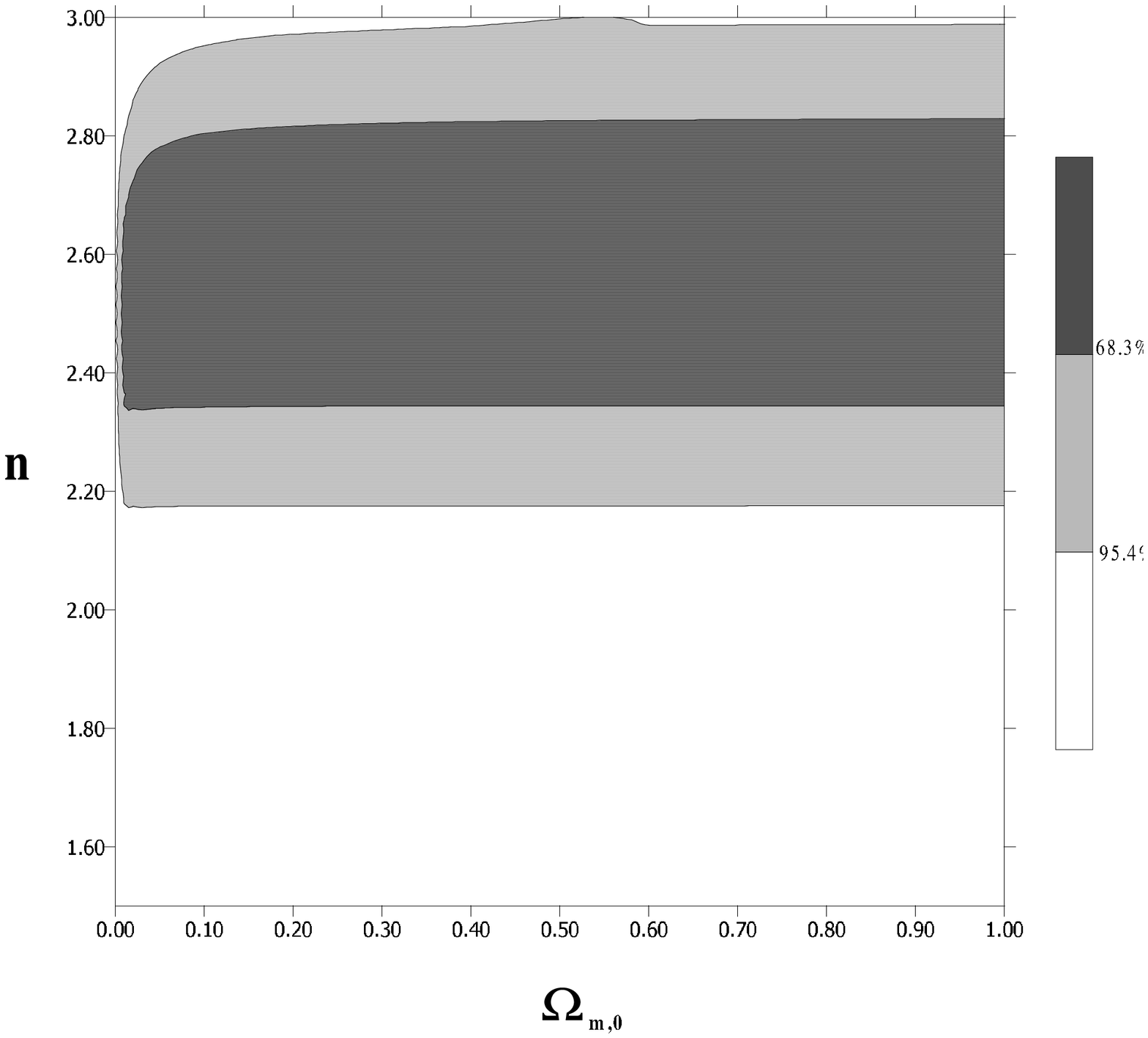}
\caption{The flat non-linear gravity model with dust and radiation
($n<3$). Likelihood contours on the $(\Omega_{m,0},n)$ plane,
marginalized over ${\cal M}$, obtained from fits to SNIa Astier et
al. sample.} \label{fig:2}
\end{figure*}

Most popular are the Akaike information criteria (AIC) \cite{Akaike:1974} and
the Bayesian information criteria (BIC) \cite{Schwarz:1978}. We use this
criteria to select model parameters providing the preferred fit to data.

One of the important problem of  modern observational cosmology is
the so-called degeneracy problem: many models with dramatically
different scenarios agree with the present day observational data.
Information criteria for model selection \cite{Liddle:2004nh} can be used,
in some subclass of dark energy models, in order to overcome this degeneracy
\cite{Godlowski05,Szydlowski06}.
Most popular are the Akaike information criteria (AIC) \cite{Akaike:1974} and
the Bayesian information criteria (BIC) \cite{Schwarz:1978}. We use this
criteria to select model parameters providing the preferred fit to data.

The AIC \cite{Akaike:1974} is defined by
\begin{equation}
\label{eq:111}
\mathrm{AIC} = - 2\ln{\mathcal{L}} + 2d \ ,
\end{equation}
where $\mathcal{L}$ is the maximum likelihood and $d$ the number of
model parameters. The best model, with a parameter set providing the preferred
fit to the data, is that which minimizes the AIC.

The BIC introduced by Schwarzc \cite{Schwarz:1978} is defined as
\begin{equation}
\label{eq:112}
\mathrm{BIC} = - 2\ln{\mathcal{L}} + d\ln{N} \ ,
\end{equation}
where $N$ is the number of data points used in the fit. While AIC tends to
favor models with a large number of parameters, the BIC
penalizes them more strongly, so the later provides a useful approximation
to the full evidence in the case of no prior on the set of model parameters
\cite{Parkinson:2005}.

Please note that both values of information criteria have no absolute sense and
only the relative values between different models are physically interesting.
For the BIC a difference of $2$ is treated as a positive evidence
($6$ as a strong evidence) against the model with larger value of BIC
\cite{Jeffreys:1961,Mukherjee:1998wp}.
If we do not find any
positive evidence from information criteria, the models are treated as
identical, while eventually additional parameters are treated as not
significant. Therefore, the information criteria offer a possibility to
introduce a relation of weak ordering among considered models.

\begin{table}
\caption{Results of AIC and BIC  performed on
the Astier versus the Gold Riess samples of SNIa.}
\label{tab:3}
\begin{tabular}{@{}p{3.5cm}rr}
\hline
sample &  AIC & BIC \\
\hline \hline
%\startdata
$\Lambda$CDM  Gold     & 179.9 & 186.0 \\
$\Lambda$CDM  Astier   & 111.8 & 117.3 \\
Non-Lin.Grav. Gold     & 186.6 & 195.8 \\
Non-Lin.Grav. Astier   & 114.7 & 122.9 \\
\hline
%\enddata
\end{tabular}
\end{table}

In the Table \ref{tab:3} the value  of  AIC and BIC for  the
$\Lambda$CDM and the non-linear gravity models are presented. Note
that for both samples we obtain  with AIC and BIC for the
$\Lambda$CDM model smaller values than for non-linear gravity. Most
interst is using a Bayesian framework to compare the cosmological
models, because they automatically penalize models with more
parameters to fit the data. Based on these simple information
criteria, we find that the SNIa data still favor the $\Lambda$CDM
model, because under a similar quality of the fit for both models,
the $\Lambda$CDM  contains less parameters.

\subsection{CMB shift parameter}

For stringent and deeper constraint on model parameters we include in our
analysis  the so called (CMB) ''shift parameter``. This parameter is defined as:
\begin{equation}
\label{eq:11}
R \equiv \sqrt{\frac{\Omega_{m,0}}{|\Omega_{k,0}|}}\mathcal{F} \left(y(z_{lss})\right)
\end{equation}
where $R_0=1.716 \pm 0.062$ \cite{Wang04} and $z_{lss}=1089$
\cite{Spergel:2003}. The $R$-parameter determines the angular scale
of the first acustic peak through the angular distance to last
scattering and physical scale of the sound horizon. It is
insensitive with respect perturbations and are suitable to constrain
model parameter. The region allowed by the analysis of (CMB) ''shift
parameter`` the plane $(\Omega_{m,0},n)$ for non-linear gravity
model (for the case $n<3$) is presented on the Fig 4 (lower panel).

We obtain for non-linear gravity model the values of the model
parameters $\Omega_{m,0}=0.67$ and $n=1.03$ as a best fit.
Please note that this area is not allowed by SNIa data.

\subsection{Baryon oscillation peak}

Recently Fairbarn and Goobar \cite{Fairbairn:2005} used baryon
oscillation peak detected in the SDSS Luminosity Red Galaxies Survey
\cite{Eisenstein:2005} as a independent test of Dvali-Gabadadze-Porrati
(DGP) brane model. They used constraint for:
\begin{equation}
\label{eq:12}
A \equiv =\frac{\sqrt{\Omega_{m,0}}}{E(z_1)^{\frac{1}{3}}}
\left(\frac{1}{z_1\sqrt{|\Omega_{k,0}|}}
\mathcal{F} \left( \sqrt{|\Omega_{k,0}|} \int_0^{z_1} \frac{d z}{E(z)} \right)
\right)^{\frac{2}{3}} \ ,
\end{equation}
so that $E(z) \equiv H(z)/H_0$ and $z_1=0.35$  yield $A=0.469 \pm 0.017$.
The quoted uncertainty corresponds to one standard deviation, where a Gaussian
probability distribution has been assumed.
These constraints  could be used for fitting cosmological parameters
(see also \cite{Astier:2005,Fairbairn:2005,Guo06}).

Fairbarn and Goobar \cite{Fairbairn:2005} showed that the joint constraints
for both SNIA data and the baryon oscillations peak ruled out flat DGP model
at the 99\% confidence level. Analogical analysis can be performed for our
model. We obtain for non-linear gravity model the values of the model
parameters
$\Omega_{m,0}=0.28$, $\Omega_{nonl,0}=0.33$ and $n=2.53$ as a best fit.
On the Fig.3 we show the region allowed by the baryon oscillation test
on the plane $(\Omega_{m,0},n)$ for non-linear gravity model
with dust and radiation (for the case $n<3$).

\begin{figure*}[ht!]
\includegraphics[width=0.9\textwidth]{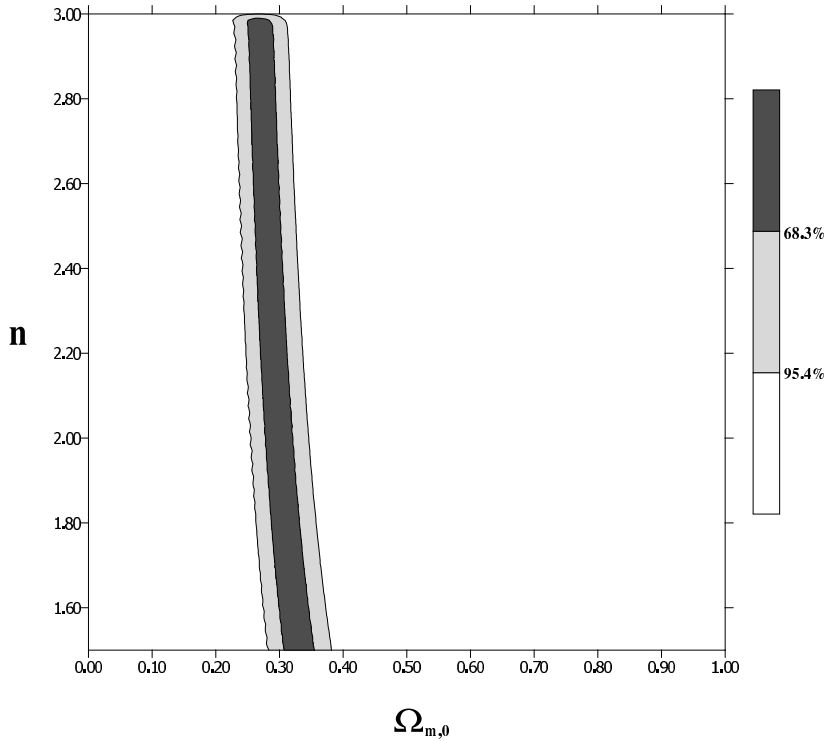}
\caption{The flat non-linear gravity model with dust and radiation
($n<3$). Likelihood contours on the $(\Omega_{m,0},n)$ plane,
marginalized over ${\cal M}$, obtained from baryon
oscillation peak test.} \label{fig:3}
\end{figure*}

\subsection{Combined SNIa, CMB shift and baryon oscillation  constraints}

Now it is possible obtained constraints from both SNIa data and CMB
shift and baryon oscillation peak. The results of our combined analysis 
are presented in the Fig.4 (upper panel) On can see that the combination
of three independent observational constraints distinguish the value
$\Omega_{m,0} \simeq 0.3$ like for $\Lambda$CDM concordance model in
which there is present substantial conception of dark matter.
However please note that area allowed by CMB shift is excluded by
area allowed by combined SNIa data and baryon oscillation peak
because different value of $n$ obtained in both cases.

\begin{figure*}[ht!]
%\centerline{\psfig{file=fig4a.eps,width=4.7in}}\\
%\centerline{\psfig{file=fig4b.eps,width=4.7in}}
\vspace*{22pt}
\includegraphics[width=0.9\textwidth]{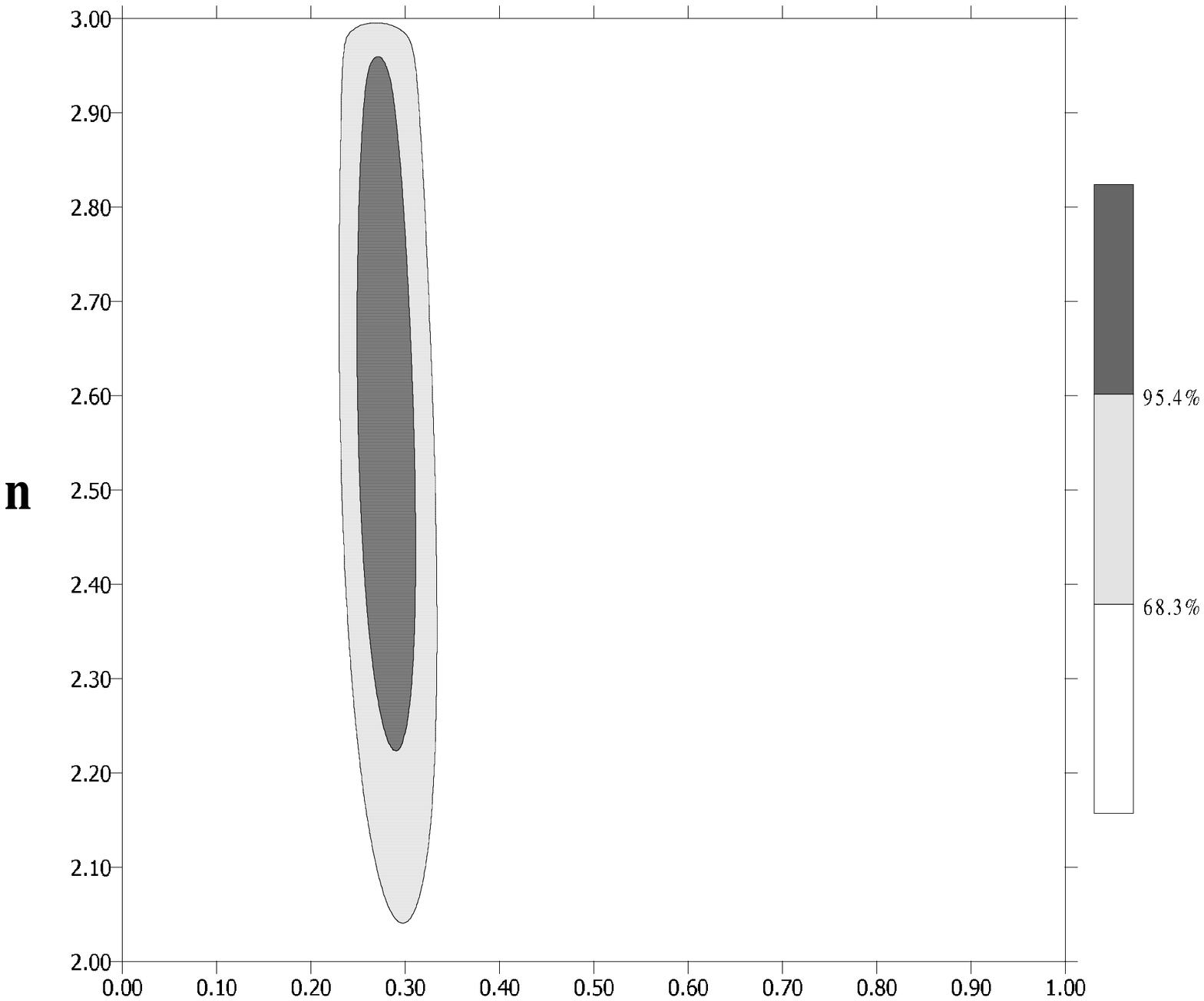}\\
\includegraphics[width=0.9\textwidth]{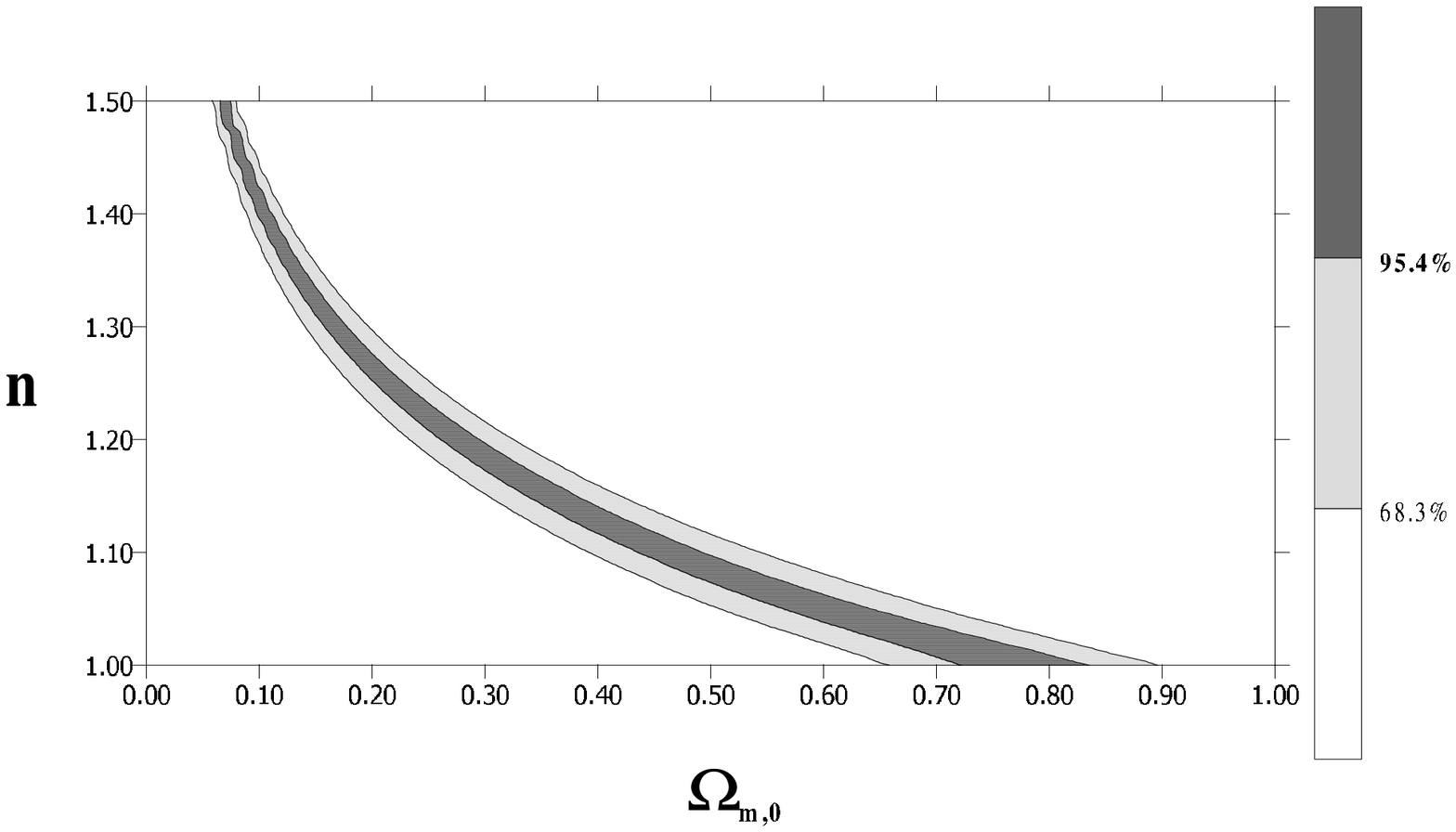}
\caption{ The comparision the the confidence levels (on the plane
$(\Omega_{m,0},n)$) obtained from combined analysis SNIa Astier
sample and baryon oscillations peak (upper panel) and from (CMB)
''shift parameter`` (lower panel) .} \label{fig:7}
\end{figure*}

\section{Flat rotation curves from cosmology in
${\cal L}\propto R^n$ theories}

Let us consider low energy limit of modified gravity with lagrangian
${\cal L}\propto R^n$. For this aims it is useful to consider point like
$m$ in Schwarzchild - like metric (spherically symmetric). Then modified
gravitational potential which corrected the ordinary Newtonian potential
is of the form:
\begin{equation}
\label{eq:1}
V(r) \propto -\frac{1}{r}(1+(\frac{r}{r_c})^{\beta})
\end{equation}
where $r_c$ is characteristic parameter which crucially depends on the
mass of the system and $\beta=\beta(n)$ \cite{Cap06a}.

Hence we can evaluate the rotation curve in the Newtonian limit of
modified gravity $\frac{mv_c^2(r)}{r}=- \frac{\partial V}{\partial
r}$. In the previous section we estimate value of $n \simeq 2.6$.
Then we calculate $\beta$ parameter from the formula:
\begin{equation}
\label{eq:2}
\beta = \frac{12n^2-7n -1 - \sqrt{36n^4+12n^3-83n^2+50n+1}}{6n^2+4n-2}
\end{equation}
obtained by  Capozzielo et al. \cite{Cap06a}.

We obtain $\beta \simeq 0.7$ which is close to $\beta$ estimated for
NGC 5023 ($\beta=0.714$). Therefore we obtain that considered theory
reproduce flat rotation curves of spiral galaxies. 
Moreover, the value of $\beta$ parameter required to explain acceleration
expansion of the Universe give rise to correct peculiarities of
observed rotation curve. Nevertheless note, that from investigation
presented in previous section density parameter for matter is close
to $\Omega_{m,0}= 0.3$ rather than to value $\Omega_{m,0}= 0.05$ as
can be expected if both effects of dark energy and dark matter has
non-substantial nature i.e. (arises from modified gravity only).

\section{Conclusion}

In this paper we consider the simplest choice of $f(R)$ theories with
$f(R) \propto R^n$. The basic motivation  is
searching for fundamental theory of gravity capable to explain both dark
energy and dark matter problems without referring to mysterious dark
energy conception. For this aim we consider cosmology based on such
a theory of gravity and then we use  different observational
constraints on independent model parameters. We consider simple flat
FRW model. It is integrable in exact form after re-parametrization
of time variable. From  estimation based on SNIa and BOP we obtain
$n>2$ which means that bouncing phase instead of big-bang
singularity is generic features of such models. Because $n>2$ new
$\tau$ parameter is monotonous function of cosmic time and
acceleration epoch is transitional only phenomenon. In the future
the universe decelerate which distinguish our model from
$\Lambda$CDM one. Note that because for small value of scale factor
a curvature effects are negligible in the comparison to other matter
contribution, therefore, in the generic case big-bang singularity is
replaced by bounce.

Analysis of SNIa Astier data shows that values of $\chi^2$ statistic
are comparable for both $\Lambda$CDM and best fitted non-linear
gravity model. For deeper analysis we use Akaike and Bayesian
information criteria of model comparison and selection. We find
these criteria still to favor the $\Lambda$CDM model over non-linear
gravity, because (under the similar quality of the fit for both
models) the $\Lambda$CDM model contains one parameter less.

Moreover, we find that the effect of dark matter can be kinematically
explained as a effect of nonlinear gravity with Lagrangian ${\cal L}
\propto R^n$. Parameter $\beta$  required for explaining accelerated
expansion of the universe give rise to  correct peculiarities of
observed rotation curve. However from baryon oscillation peak prior
we still obtain $\Omega_{m,0} \simeq 0.3$ (instead of $\Omega_{m,0} \simeq
0.05$ as we expected). Moreover, we find a disagreement between results
obtained from CMB shift parameter analysis and that from joint SNIa
and baryon oscillation peak. Finally, the substantial form of dark matter
is still required.

\end{document}